# Generation of High-Purity Millimeter-Wave Orbital Angular Momentum Modes Using Horn Antenna: Theory and Implementation

Jian Ren[1,2*], Kwok Wa Leung[1,2]

[1] Department of Electronic Engineering, City University of Hong Kong, Kowloon, Hong Kong SAR, China

[2] State Key Laboratory of Millimeter Waves, City University of Hong Kong, Kowloon, Hong Kong SAR, China

*correspondence to jianren2-c@my.cityu.edu.hk

**Abstract**

Twisted electromagnetic waves, of which the helical phase front is called orbital angular momentum (OAM), have been recently explored for quantum information, high speed communication and radar detections. In this context, generation of high purity waves carrying OAM is of great significance and challenge from low frequency band to optical area. Here, a novel strategy of mode combination method is proposed to generate twisted waves with arbitrary order of OAM index. The higher order mode of a circular horn antenna is used to generate the twisted waves with quite high purity. The proposed strategy is verified with theoretical analysis, numerical simulation and experiments. A circular horn antenna operating at millimeter wave band is designed, fabricated, and measured. Two twisted waves with OAM index of l=+1 and l=-1 with a mode purity as high as 87% are obtained. Compared with the other OAM antennas, the antenna proposed here owns a high antenna gain (over 12 dBi) and wide operating bandwidth (over 15%). The high mode purity, high antenna gain and wide operating band make the antenna suitable for the twisted-wave applications, not only in the microwave and millimeter wave band, but also in the terahertz band.

# Introduction

Exploiting all aspects of electromagnetic (EM) wave properties has been quite a hot topic in the past several decades, from low frequency radio waves to optical waves. EM wave has brought big changes to our daily lives. In the meanwhile, it also attracts a wide range research on exploring its new characteristics, including, but not limited to wavelength, polarization and amplitude. Among these properties, orbital angular momentum (OAM) [1, 2], featuring a helical phase front plane, is a newborn, compared with the spin angular momentum (SAM) [3]. As been known long before, EM waves could carry both OAM and SAM [3, 4]. However, the first detailed research on the OAM was carried out by Allen in 1992 [1], who showed that a light with a Laguerre-Gaussian amplitude distribution have an OAM of $l\hbar$ per photon, where $l$ is the OAM mode index and $\hbar$ is the Planck's constant. Since then, how to analyze, generate, and utilize the EM waves carrying OAM attracts researchers of various academic areas all over the world.

In the quantum area, transferring from SAM to OAM of photons was used to transpose the quantum information [5]. With the scattering of an OAM light from a spinning object, the rotate speed of the rotating bodies can be detected, using the characteristic of OAM combining Doppler effects [6]. This can also be used in astrophysics to detect the running state of the celestial body. In the wireless communication area, OAM waves find the most attractive and potential applications, since they can enhance the data capacity of the communication systems as the natural orthogonality between different OAM modes [7]. This idea was demonstrated in the free space information transfer in 2004 for the first time [8]. Thereafter, terabit data transmission has been realized using this concept in the free space in 2012 [9] and in fiber in 2013 [10]. This breakthrough happens not only in the optical region, but also in the microwave and millimeter wave bands [11, 12].

Although different methods have been proposed to generate OAM, the generation of high purity OAM is still a big challenge. The spiral phase plate (SPP) [13] is most often used to create helical phase structure in optical region. When the beam penetrates into the SPP, a phase difference can be generated along the azimuthal angle due to the variation of the SPP thickness, making the traditional plane-wave wave front a helical one along the direction of the wave propagation. This method has an advantage of high precision, which can generate OAM beam with a mode index as high as 5050 [14]. Recently, metasurfaces [15-17] also have been used to generate OAM beams [18-20]. The working principle is somewhat similar with that of the SPP, as the element phases reflected/transmitted by the metasurface have the same phase variation. Another method to create OAM beam is resonator cavity. Using a modified micro-ring resonator, Cai [21] demonstrated a very compact OAM beam emitter on a single chip. At the microwave and millimeter-wave bands, the first OAM beam generation is demonstrated by Thide in 2007 with circular distributed antenna array [22]. The circularly-placed antenna elements have a phase variation along the azimuthal direction. Antenna Array can generate an OAM beam with mode index $|l|<N/2$, where $N$ is the number of antenna elements on a circle around the beam axis. Since then, many antenna arrays radiating the OAM beams at microwave band were investigated and related communication experiments were reported [23-25]. In addition, this method has also been employed for optical phase array at optical frequency [26-28]. Although all the methods mentioned above can generate OAM beams, the realization is complicated and space-consuming. To address the issue, generating EM waves carrying OAM with only one single radiator becomes a hot topic recently. These attempts include traveling wave loop antenna [29, 30], and patch antenna [31] and reflector antenna [32].

Horn antenna [33-36] is one of the most widely used antennas in the wireless region, from communication to detection, from military to commercial applications. The horn antenna owns advantages of easy fabrication, high gain, high power handling, and stable phase center, compared to other types of antenna. However, until now, the horn antenna can radiate OAM wave only with metasurface or SPP mentioned above, resulting in the big volume of the whole antenna structures.

In this paper, we propose a new strategy to generate EM wave with OAM using the horn antenna. This strategy utilizes the higher order mode of the circular horn antenna, which has been overlooked for a long time. With the excitation of appropriate higher order mode, OAM wave with arbitrary charge can be generated. We gave the computational formulae of the proposed antenna's radiation field, which theoretically prove that the field carries OAM. With full-wave simulation software, we simulated the near and far fields of the proposed horn antenna, which agree well with the theoretical ones. Finally, a prototype of the antenna working at millimeter wave band is fabricated and measured. The measured results coincide with the simulated ones, further verifying the design philosophy. To the best of the authors' knowledge, it is the first time to generate OAM wave by a single horn antenna without extra component. It should be noted that the strategy is applicable not only to the millimeter waves but also to other electromagnetic spectrum, e.g. the terahertz wave, which is a quite emerging topic nowadays.

**Results**

### I. Design Principle and Theoretical Analyses

The schematic diagram of the designed horn antenna radiating EM waves carrying OAM (twisted waves) is shown in Figure 1 (a). The antenna consists of two rectangular waveguide feeding ports, a circular waveguide and the tapered opening of the circular waveguide. The two feed ports are excited with two signals with a same amplitude and a phase difference of +90º or −90º as shown in Figure 1(a). The angle between the two feeding ports is $\alpha$, which can be determined according to the required OAM mode index $l$. Figure 1(b) gives the schematic diagram of the coupler, which is used to provide the desired amplitude and phase of excitation for the horn antenna . In practice, the two output ports are connected to the two feed ports of the horn antenna. The outputs of the coupler (indicated as outport1 and outport2) have an equal amplitude but a 90º phase difference. The symbol of the phase difference is determined by the input. When the port 1 is excited, the phase difference between the outport1 and putport2 is +90° and an EM wave carrying OAM of index of $l$=+1 can be generated.

Likewise, with the port 2 excited, an OAM wave with index of *l*=−1 can be obtained due to the −90° phase difference. A prototype of the proposed horn antenna then designed and fabricated, as shown in Figure 1(c). The two feeding ports of the horn antenna are connected to the output ports of a coupler via a coaxial-waveguide adapter and coaxial cable.

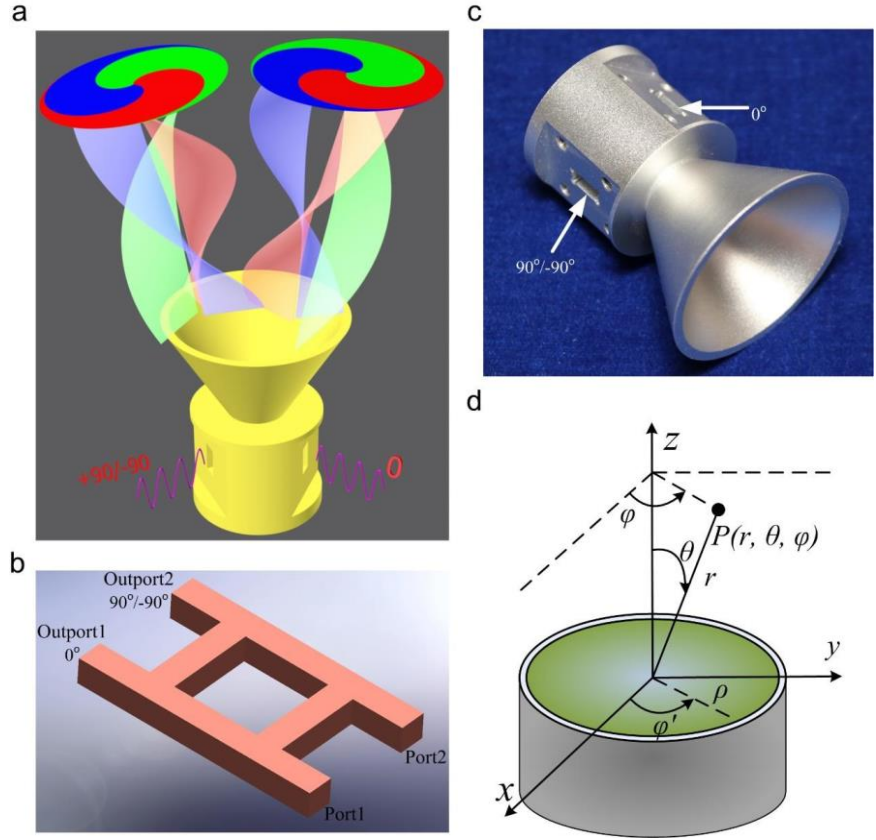

Figure1 | Schematic diagram of the designed horn antenna radiating electromagnetic wave carrying OAM. (a) Schematic diagram of the designed horn antenna. The two ports are excited with microwave signals with a phase difference of +90° or -90°, and EM waves carrying OAM with index of l=+1 and l=-1 can be therefore generated. (b) Schematic diagram of a 90° hybrid coupler. The two output ports have a phase difference of +90° or -90° determined by Port1 or Port2 excitation. (c) Prototype of the designed horn antenna without feed network. (4) Cylindrical coordinate system of the circular aperture in the free space.

To better understand the OAM principle of the antenna, the formulae of the radiation filed of the designed horn antenna is needed. The computational formulae of a circular aperture's radiation field [37], expressed in the Cartesian coordinate system as shown in Figure 1 (d), are derived to estimate the radiation characteristic of a conical horn antenna with a circular waveguide. Sharing the same original point of the formulae in [37], a spherical coordinate (r, ϑ, φ) is used to locate arbitrary point P in the free space. In this paper, only TE mode is considered for OAM excitation. According to the cavity model in cylindrical coordinates (see Figure 1(d)), the guided electric field dominated by a TE$_{mn}$ mode in the circular waveguide can be expressed in the cylindrical coordinates as:

$$E_\theta = j\omega\mu k_{mn} cos(m\varphi')J_m'(k_{mn}\rho) \times e^{-j\beta z};$$

$$E_\rho = j\omega\mu m sin(m\varphi')\frac{J_m(k_{mn}\rho)}{\rho} \times e^{-j\beta z}; \qquad (1)$$

$$E_z = 0;$$

where *m* and *n* are the mode indices in azimuthal and radial direction, respectively. $J_m$ is the first kind Bessel function with an order *m* while $J_m'$ donates the derivative of $J_m$. The radiation fields can be then deduced as follows,

$$E_\theta = j^{m+1}\frac{m\omega\mu}{2r}\left[1+\frac{\beta_{mn}}{k}\cos\theta+\Gamma(1-\frac{\beta_{mn}}{k}\cos\theta)\right]J_m(k_{mn}a)\frac{J_m(k_0 a\sin\theta)}{\sin\theta}\sin(m\varphi)e^{-jk_0 r},$$

$$E_\varphi = j^{m+1}\frac{k_0 a\omega\mu}{2r}\left[\frac{\beta_{mn}}{k}+\cos\theta-\Gamma(\frac{\beta_{mn}}{k}-\cos\theta)\right]J_m(k_{mn}a)\frac{J_m'(k_0 a\sin\theta)}{1-\left(\frac{k_0\sin\theta}{k_{mn}}\right)^2}\cos(m\varphi)e^{-jk_0 r}, \quad (2)$$

$$E_r = 0,$$

where $\Gamma$ is the reflection coefficient in the waveguide, which can be regarded as zero under ideal condition. That is to say, the waveguide is ideally matched. When the reflection coefficient is ignored, Eq. (2) can be simplified as:

$$E_\theta = Am\left[1+\frac{\beta_{mn}}{k}\cos\theta\right]\frac{J_m(k_0 a\sin\theta)}{\sin\theta}\sin(m\varphi),$$

$$E_\varphi = Ak_0\left[\frac{\beta_{mn}}{k}+\cos\theta\right]\frac{J_m'(k_0 a\sin\theta)}{1-\left(\frac{k_0\sin\theta}{k_{mn}}\right)^2}\cos(m\varphi), \quad (3)$$

$$E_r = 0,$$

where $A = j^{m+1}\frac{\omega\mu}{2r}J_m(k_{mn}a)e^{-jk_0 r}$.

Given the known field distribution of $TE_{mn}$ mode, the two input signals are considered subsequently. Here, the two feeding ports, with a rotation angle of *α*, are driven with the signals of same amplitude but orthogonal phase. The value of *α* is the angular spacing of the probe depending on the mode order and can be expressed as *α* = (2×b+1)π/2m, where b is an integer [38]. Considering the phase difference between port1 and port2 is positive 90 degree, which means the port 2 have a 90 °phase delay to port 1, the total field can be written as follows, after the superposition of the individual electric fields excited by the two orthogonal modes,

$$E_\theta = E_\theta^1(r,\theta,\varphi) + E_\theta^2(r,\theta,\varphi) = E_\theta^1(r,\theta,\varphi) + jE_\theta^1(r,\theta,\varphi+\alpha)$$
$$= jAm[1+\frac{\beta_{mn}}{k}\cos\theta]\frac{J_m(k_{mn}a)}{\sin\theta}e^{-jm\varphi},$$

$$E_\varphi = E_\varphi^1(r,\theta,\varphi) + E_\varphi^2(r,\theta,\varphi) = E_\varphi^1(r,\theta,\varphi) + jE_\varphi^1(r,\theta,\varphi+\alpha)$$

$$= jAm\left[1+\frac{\beta_{mn}}{k}\cos\theta\right]\frac{J_m(k_{mn}a)}{\sin\theta}e^{-jm\varphi} = Aka\left[\frac{\beta_{mn}}{k}+\cos\theta\right]\frac{J_m'(k_0 a\sin\theta)}{1-\left(\frac{k_0\sin\theta}{k_{mn}}\right)^2}e^{-jm\varphi} \quad (4)$$

$$= \frac{1}{2}Aka\left[\frac{\beta_{mn}}{k}+\cos\theta\right]\frac{J_{m-1}-J_{m+1}}{1-\left(\frac{k_0\sin\theta}{k_{mn}}\right)^2}e^{-jm\varphi},$$

$$E_r = 0,$$

where the radiated fields of the aperture from the two input signals can be donated as $E_\theta^{\ 1}(r,\theta,\varphi)$ and $E_\theta^{\ 2}(r,\theta,\varphi)$. Using coordinate transformation between spherical coordinates and Cartesian coordinates, the field distribution at arbitrary point can be derived,

$$E_x = E_\theta \cos\theta \cos\varphi + E_\varphi(-\sin\varphi)$$

$$= jB\left(\cos\theta + \frac{\beta_{mn}}{k}\cos^2\theta\right)(J_{m-1} + J_{m+1})e^{-jm\varphi}\cos\varphi$$

$$- B\left(\cos\theta + \frac{\beta_{mn}}{k}\right)\frac{J_{m-1} - J_{m+1}}{1 - \left(\frac{k_0\sin\theta}{k_{mn}}\right)^2}e^{-jm\varphi}\sin\varphi,$$

$$E_y = E_\theta \cos\theta \sin\varphi + E_\varphi(\cos\varphi)$$

$$= B\left(\cos\theta + \frac{\beta_{mn}}{k}\cos^2\theta\right)(J_{m-1} + J_{m+1})e^{-jm\varphi}\sin\varphi \qquad (5)$$

$$- B\left(\cos\theta + \frac{\beta_{mn}}{k}\right)\frac{J_{m-1} - J_{m+1}}{1 - \left(\frac{k_0\sin\theta}{k_{mn}}\right)^2}e^{-jm\varphi}\cos\varphi,$$

$$E_z = 0,$$

Where $B = \frac{1}{2}Aka$.

Since only the value of OAM along the wave propagation direction is taken into consideration, θ is deemed as very small. Under this circumstance, cos1 and sin0 could be applied for the simplification of the formulae:

$$E_x = B\left(\cos\theta + \frac{\beta_{mn}}{k}\right)\left(J_{m+1}e^{-j(m+1)\varphi} + J_{m-1}e^{-j(m-1)\varphi}\right)$$

$$\approx BJ_{m-1}e^{-j(m-1)\varphi}\left(\cos\theta + \frac{\beta_{mn}}{k}\right),$$

$$E_y = jB\left(\cos\theta + \frac{\beta_{mn}}{k}\right)\left(J_{m+1}e^{-j(m+1)\varphi} + J_{m-1}e^{-j(m-1)\varphi}\right) \qquad (6)$$

$$\approx jBJ_{m-1}e^{-j(m-1)\varphi}\left(\cos\theta + \frac{\beta_{mn}}{k}\right),$$

$$E_z = 0.$$

From the calculated results, it can be seen that x-component and y-component of the electric fields have the azimuth dependency of $e^{-j(m-1)\varphi}$, indicating the twisted waves. When the phase difference between port 1 and port 2 is negative 90 degree, the azimuth dependency of $e^{+j(m-1)\varphi}$ can be obtained. That is to say if we want to obtain EM wave carrying OAM of |l| order using circular horn antenna, the $TE_{l+1,n}$ mode should be excited according to (6). The sign of the phase difference between the two signals can result in different sign of the OAM beam's index, e.g., the positive corresponding positive mode index and negative corresponding negative mode index.

To verify the computational formulae of the radiation fields obtained above, a Matlab program is used to clarify the exact phase and amplitude difference of the ideal case. Figure 2 (a) gives the two dimensional visualizations of the phase and amplitude patterns of the electric field components corresponding to different OAM mode indices $l$, where the circular observation window lies at $z=1000\lambda$ with a radius of $\rho=80\lambda$. With reference to the subfigure, the left half are the phase patterns when parameter $l$ varies from 0 to +2, while the right half depict the amplitudes correspondingly. The positive sign of $l$ denotes a clockwise phase helix, which can be seen in the phase pattern, and vice versa. It should be noted that here we only give the cases of positive $l$ value and those of negative $l$ can be found in Supplementary figure S2

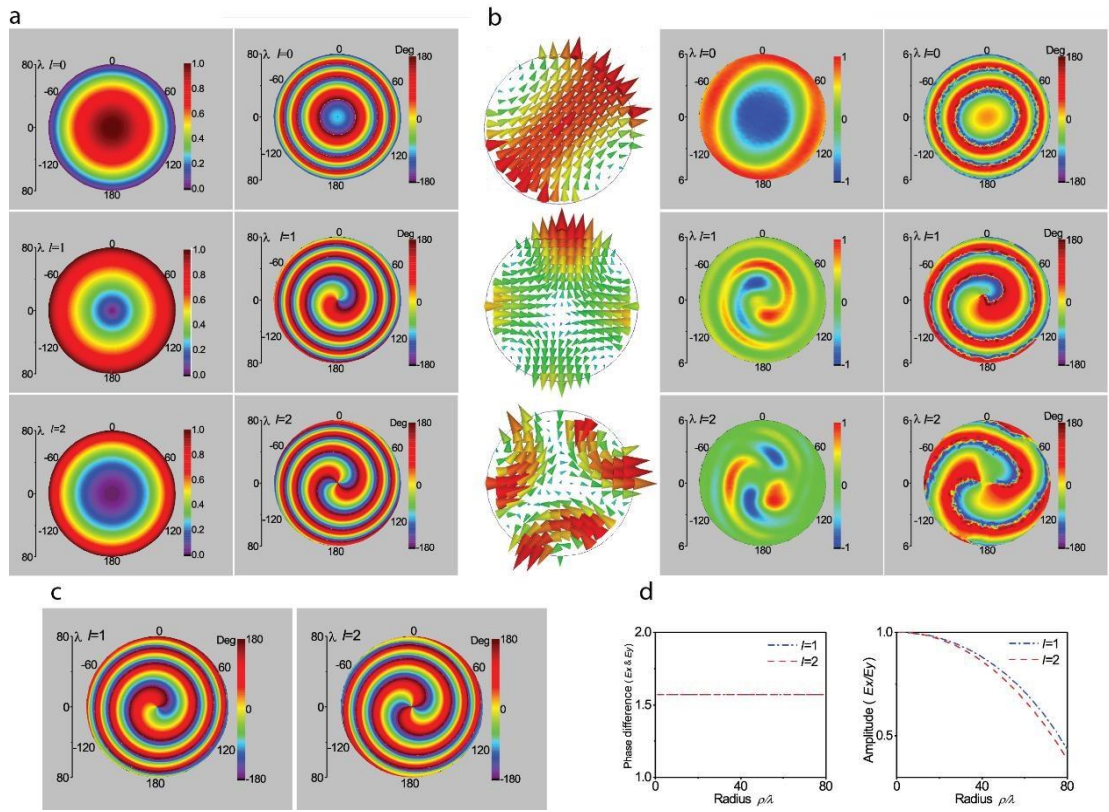

Figure2 | Calculated and simulated results of the filed intensity and phase of the field. (a) 2-D phase and amplitude patterns of the twist beams calculated by Matlab with different OAM index $l$ at $z=1000\lambda$, the circular observation window has a radius of $\rho=80\lambda$. The left half is the field intensity and the right half is the phase of the field. From top to bottom: $l=0$, $l=+1$, $l=+2$. (b) Simulated 2-D field and phase distribution using ANSYS HFSS corresponds to different mode index $l$. From left to right: the vector field distribution in the circular waveguide corresponding to different mode (From top to bottom: $TE_{11}$, $TE_{21}$, $TE_{31}$), the transient field with phase information with different mode index (From top to bottom: $l=0$, $l=+1$, $l=+2$) and phase pattern (From top to bottom: $l=0$, $l=+1$, $l=+2$). The circular observation window locates at $z=100\lambda$ with a radius of $\rho=80\lambda$ (c) Angle pattern of the transverse field with $l=+1$ and $l=+1$. The circular observation window is same with that of figure (a). (d) Ratio of the phases and amplitudes of the two field components Ex and Ey. The observation window is located at $z=1000\lambda$

With reference to Figure 2(a), one can clearly see that the phase changes of the electric field along a concentric circle, which goes consistent with the current phase variation along the wire. The rule for phase change of the radiation electric field is $2\pi$ multiplied by $l$. That is to say, when $l=+1$, the phase changes $2\pi$ during one revolution, while a $4\pi$ phase change can be observed when $l=+2$. The rest can be done in a similar manner. According to Figure 2(a), it can be found that the amplitude of the radiation in the center of the beam keeps null for the $l=+1$ and $l=+2$ mode, similar to the results observed in the twisted optic beam possessing the OAM mode. It can be also observed that the wave divergence becomes severer as the $l$ gets larger, which can be found in the twisted beam using the other types of antennas as well. It is worth noting that when $l=0$, the maximum value of the amplitude occurs at the center of the beam. This is consistent with the conventional circular horn antenna, which utilizes the $TE_{11}$ mode to generate a high gain unidirectional radiation pattern. There is no OAM mode included in the radiation beam for $l=0$ and this is the reason why there is no phase change along the azimuth angle. To see the polarization of the twisted radio waves from the direction of the transmission axis ($z$ orientated), the amplitude and the phase of the axial ratio are calculated and the results are shown in Figure 2(d), after we define the axial ratio of the two components of the field as $\xi=|E_x/E_y|$. The observation plane of Figure 2(d) has the same dimension as that of Figure 2(a). With reference to the figure, we can see that the twisted radio waves are circularly polarized along the propagation direction as the amplitude difference of $\xi$ equals to 1 and the phase difference is 1.57 ($\pi/2$) at $\rho=0$. Moreover, as the radius of the observation plane becomes larger, the amplitude difference of $\xi$ becomes smaller, implying a bad performance of circular polarization. A comparison between conditions of $l=1$ and $l=2$ is shown in Figure 2(d). As can be observed in the figure, the variation of axial ratio is faster when $l=2$, because the direction of the maximum power radiation owns a larger included angle with $z$-axis. The difference between the two conditions can be also predicted by the amplitude distributions in Fig. 2(a).

The angle between the transverse field and the radius vector in the cylindrical coordinate system can be represented by the direction of transverse field, signified by the angle $\sigma=\arctan[Re(E_x/E_y)]$. Figure 2(c) displays the calculated results under the same observation plane for easy comparison. Obviously, the number of the arm is twice the value of $l$.

To further verify our theory, the numerical simulation is carried out using the full wave simulation software ANSYS HFSS. As mentioned above, the index of OAM carried by radiation fields has a relationship with the mode index of the excited mode in circular waveguide, namely $l=m-1$. Therefore, the OAM beam with order $l$ corresponds to the $TE_{l+1, n}$ mode in the waveguide. Figure 2 (b) gives the vector field distributions of different modes in the waveguide, along with the transient field distributions and the phase patterns of the radiation electric field with corresponding order of $l$. In the simulation, two feed waveguides with a phase difference of 90° are used to excited different working modes of the circular waveguide. One thing that should be noted is that the observation window is much smaller than that in the calculation with MATLAB, as the full wave simulation software requires extensive computing resources and the model cannot be too large. The observation window with a radius of $\rho = 6\lambda$ is therefore chosen to be located at $z = 5\lambda$ and all the results are shown in Figure 2(b). From the results, it can be seen that when the $TE_{11}$ mode is excited (first row of Figure 2(b)), the corresponding index of the OAM beam is $l=0$, which can be predicted using the phase pattern and transient field distributions. This is also applicable for the $TE_{21}$ and $TE_{31}$ mode, corresponding to $l = +1$ and $l$ = +2. Two arms for l = +1, and four for l = +2 (the red and blue arms in the figure) in the transient field distributions can be seen apparently, illustrating the existence and the index of OAM mode, while no arm can be observed in the transient field distribution which corresponds to $l$=0 mode. A similar phenomenon also can be observed in the phase patterns shown in the third column of Figure 2(b). For the phase patterns with $l$ = +1 and $l$ = +2, one arm and two arms can be in turn found along the counterclockwise direction, while the field and phase keep constant along the azimuth angle for $l$ = 0 case and no arm can be observed. In addition, from the results shown in Figure 2, very good agreement can be found between the simulated phase patterns and the calculated ones, having verified the theory.

**II. Experimental verification and results**

To verify the theory and design principle analyzed above, a conical circular horn antenna carrying OAM with order $l$=±1 at millimeter wave band is fabricated and measured. The photos of the fabricated horn antenna are shown in Figure 1 (c). Figures 3(a) and 3(b) give the simulated and measured $S$ parameters of proposed antenna, respectively. Excellent agreement between the simulation and measurement can be observed. As indicated in Figure 3(a), the 10-dB impedance bandwidths of port 1 and port 2 can fully cover 32–40 GHz. The simulated and measured $S_{21}$ (mutual coupling) between the two input ports are compared in Figure 3(b). It can be seen that the measured isolation is higher than 15 dB throughout 31–37 GHz. Considering the overlapping bandwidth, the operation bandwidth is about 15% (32–37 GHz), which is much wider than other results in the open literature. A frequency shift about 1% between the simulated and measured results can also be found, owing to the fabrication error primarily.

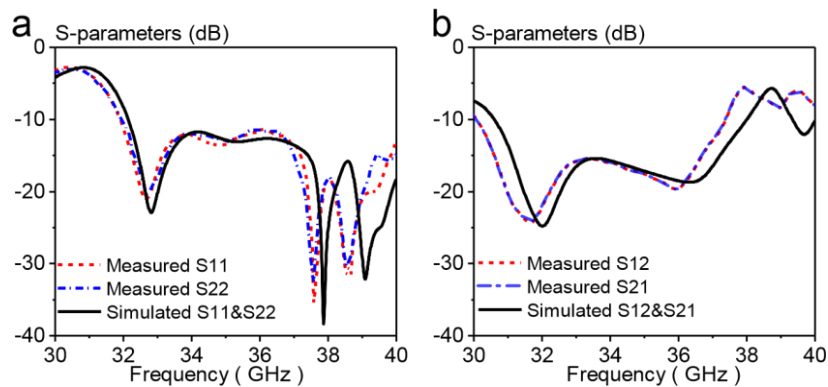

Figure 3 | Simulated and measured $S$-parameters of proposed antenna (a) Simulated and measured $|S_{11}|$ and $|S_{22}|$. (b) Simulated and measured $|S_{21}|$ and $|S_{12}|$. The solid black lines indicate the simulated results and the blue and red dotted lines indicate the measured results.

Using the near-filed antenna measurement system provided by NSI cooperation, the near filed phase pattern and the power density were measured, including both the cases of $l$=+1 and $l$=-1. The details about the measurement system and setup are available in the method section. To give a more intuitive cognition on the generated twisted wave, the field intensity and phase pattern of the radiation field on the planes with different distance are shown. Figures 4 (a)–(d) give the measured and simulated field intensity and the phase pattern at $z$ = 85 mm (z = 10 $\lambda$) and z = 400 mm (z = 50 $\lambda$). The area is a rectangle with dimensions of 200mm×200mm. From the field intensity, a power null can be obviously observed at the center of the scanned area. The difference between the maximum and minimum values is more than 25dB, which can be observed in the field intensity pattern at different distances z. For the phase pattern, a clear anticlockwise helical phase structure can be found for the $l$=+1 mode. In this case, the phase change along one circle is 2π which is also consistent with the calculated results. On the other hand, for the case of $l$ = -1, a similar phase structure can be observed while the phase change along the azimuth angle is anticlockwise. Compared with the phase pattern at z = 400 mm, the phase pattern at z=85mm has more turns in the same scanned area. From the figure of the field intensity at z= 400 mm (Figure 2(c) and (d)), we can see that the maximum power occurs at the $\rho$ = 30 mm (4$\lambda$). According to the communication principle of the OAM, the best communication link can be obtained if the maximum power area is involved at the receiving antenna. Therefore, if the distance between two OAM antennas is far, the required apertures of the antennas will big, making the system bulky. That is why the OAM is limited in the near field area in the wireless communication systems.

With the information of the field intensity and phase pattern, a quantitative research on the weight of the generated OAM beam (topological charge) is carried out by using the spiral spectrum algorithm [39-41]. The method is quite similar to a Fourier transform and it can be used to extract the weight of different spiral harmonics of EM field. Calculated from the phase distribution with this method, the extracted topological charge spectrum from $l$=-3 to $l$=+3 is shown in Figure 4 (e). With reference to the figure, one can see that for $l$=+1, over 87% of the total power is carried by the azimuthal component $e^{i\varphi}$ in the simulated and measured results. For $l$=-1 case, the number is 89.7%. This is to say, for the two cases, most of the power is carried by the fundamental azimuthal component and the generated OAM beam has a topological charge with high purity.

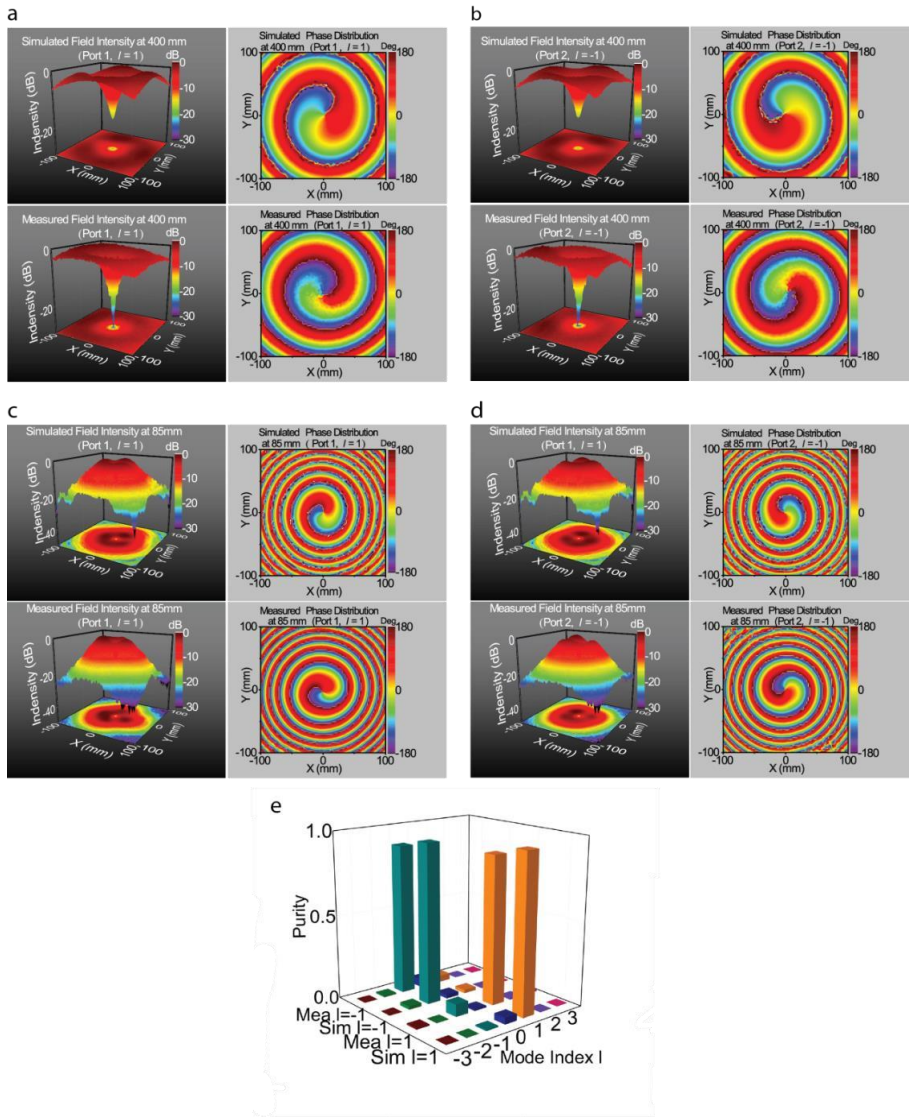

Figure 4 | Simulated and Measured Field intensity and phase pattern (a) Simulated and measured field intensity and phase distribution for mode $l$=1. The observation window is located a distance of z=400mm (50$\lambda$). (b) Simulated and measured field intensity and phase distribution for mode $l$=-1. The observation window is located a distance of z=400mm (50$\lambda$). (c) Simulated and measured field intensity and phase distribution for mode $l$=1. The observation window is located a distance of z=85mm (10$\lambda$). (b) Simulated and measured field intensity and phase distribution for mode $l$=-1. The observation window is located a distance of z=85mm (10$\lambda$). The scanned area in figure (a-d) is 200mm×200mm and all of the values are normalized. The left half of figures is field intensity and the right half is the phase pattern. The top half of the figures is simulated results while the bottom half is measured results. (e) Simulated and measured mode purity using the spiral spectrum algorithm.

Using the same near filed measurement system and the near field to far field transformation, the far field characteristics of the horn antenna is measured as well, which have been rarely involved in the open literatures. Figure 5 (a) shows the simulated and measured 3D far field radiation patterns of the proposed antenna. Since the scan plane is limited (see method section), we only give the radiation pattern in the range of -60°< $\theta$ < 60°. In order to have a more deep understanding on the antenna radiation fields, the 2D radiation patterns in different planes of the antenna are also shown in Figure 5 (b) at the same time, where good agreement between the measured and simulated far field of the antenna can be observed. All the results are normalized and the coordinates are defined in accordance with Figure 1. The cases of *l*=+1 and *l*=-1 are included, corresponding to port 1 and port 2, respectively. From the results, it can be found the maximum radiation is in the θ=12 °direction and a clear deep null can be found at z axis ($\theta=0$), indicating that the power density is zero along the propagation axis. From the 2D radiation pattern, we can see that the field intensity in the central region of the vortex is about 23 dB lower than that in the maximum intensity region. The measured gain and efficiency of the horn antenna are shown in Figures 5 (c) and (d), respectively. The measured peak gain of the antenna is 12.7dBi over the whole operation band, relatively high compared with some other type antennas. In the whole operation band, the antenna efficiency is higher than 80%. However，compared with the commercial horn antennas, the antenna efficiency is somewhat low. This is because the coupler used has a high insertion loss (more than 1 dB), which can be improved by using a millimeter wave coupler based on the waveguide technology.

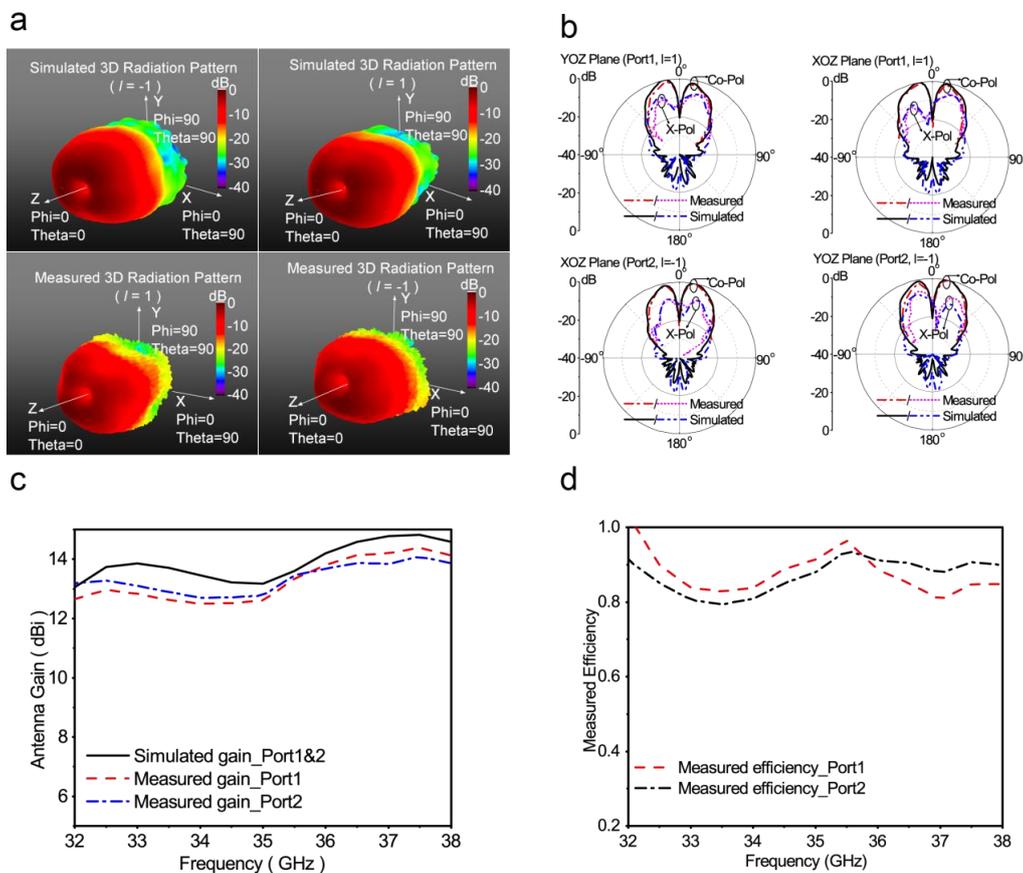

Figure 5 | Simulated and measured far field characteristics of the antenna (a) Simulated and measured 3D radiation pattern of the antenna. (b) Simulated and measured 2D radiation pattern of the antenna in different planes. All the values are normalized. (c) Simulated and measured antenna gain. The solid lines denote simulation and the dashed lines denote measurement. (d) Measured radiation efficiency of the horn antenna.

**III. Link measurement of OAM mode multiplexed horn antenna (l=+1 and l=-1)**

As mentioned in the introduction section, the helical phase structure of the EM carrying OAM has a natural orthogonality between different modes. This makes the OAM-based communication a promising technology for the next generation communication systems, from the fiber communication to wireless links. To verify the orthogonality between the l = +1 and l = -1 mode, we carry out a communication link measurement composed of these two OAM modes simultaneously. The schematic diagram of the setup is shown in Figure 6. In the experiment, two designed horn antennas, one as a transmitting antenna and the other as a receiving antenna, are placed in the z axis with concentricity and separated with a distance of 80 mm (10$\lambda$), which can be considered as a near field area. The two hybrid couplers are connected to the horn antenna via coaxial cables. During the measurement, -6 dBm transmitting power is utilized, which is 0.25 mW. The transmission coefficients ($S_{21}$) for transmitting antenna and receiving antenna were measured by network analyzer. Table I shows the measured absolute and normalized magnitudes of $S_{21}$ for the transmitting and receiving antennas. From the results, isolation over 14.5 dB were obtained between the two transmission channels, proving the orthogonality of the different modes of twist waves generated by the horn antenna.

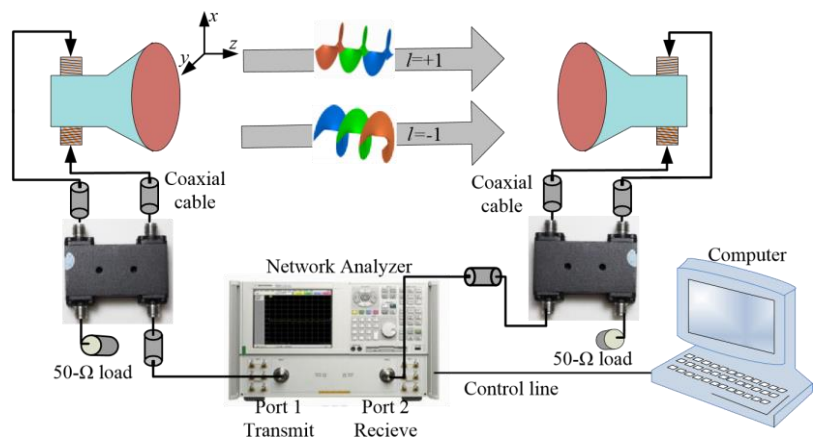

Figure 6 | Schematic diagram of the communication link measurement for two OAM modes (*l*=+ 1 and *l*=-1)

Table I Measured isolations between the different modes of twisted waves

| $S_{21}$/Normalized(dB) | $l_{receive}$ =+1 | $l_{receive}$ =-1 |
|---|---|---|
| $l_{transmit}$ =+1 | -54.5 / 0 | -69.7 / -15.2 |
| $l_{transmit}$ =-1 | -68.6 / -14.5 | -54.1 / 0 |

## Discussion

Here, a novel strategy generating twisted waves with arbitrary order OAM mode using horn antenna is proposed. For the first time, higher order modes in horn antenna, which have been overlooked for a long time, are introduced to generate twisted waves with mode combination in this new strategy. Based on the theoretical derivation of the operating principle, the characters of the generated twisted waves, such as the phase pattern, intensity distribution and polarization state, are discussed in detailed. To verify the theory, the numerical simulation of the strategy is carried out and then a prototype of the designed horn antenna which can generate the *l* = +1 and *l* = -1 OAM mode is fabricated and measured.

The fabricated prototype of the antenna, which works in millimeter wave band, can generate twisted waves carrying quite high purity OAM modes. With reference to the measured results, it can be indicated that the purity of the OAM mode is higher than 87%, both for *l* = +1 mode and *l* = -1 mode.

The antenna possesses a wide bandwidth (over 15%) and relatively high antenna gain (over 12 dBi). In addition, a communication link measurement composed of two OAM modes is carried out to verify the orthogonality between the $l$ = +1 and $l$ = -1 mode. Over 14-dB orthogonality can be observed in the results.

As one of the most widely used antenna in the electromagnetic area, horn antennas are used almost in every branch, from communication to measurement. Therefore, realizing the high purity OAM mode generation with horn antenna has great significance. This antenna's operation band can be easily shifted to Terahertz band, which is an emerging and promising area recently but little work has been done in terahertz twisted wave generation. The performance parameter, such as the bandwidth and the gain, can be further improved using the conventional optimization method in the horn antenna design. In a word, the simple working principle, superior performance and easy fabrication make the antenna promising in the future OAM waves applications.

**Method**

The calculated results in the theory parts are all processed with MATLAB software. The proposed horn antenna is designed and optimized using the commercial available full-wave simulation software ANSYS HFSS (High Frequency Structural Simulator), based on the finite element method. According to the simulations and designs, we manufactured a prototype of the horn antenna using machining manufacturing. The horn structure is made of aluminum alloy and surface anti-oxidation processing was applied. A commercially-available 90° hybrid coupler is used as feeding network. The S-parameters are measured through network analyzer 8361a provided by Keysight Technologies.

The NSI near field measurement system setup is shown in Figure 7 (a). An open waveguide is used as a probe to measure the phase and field intensity. The probe is fixed on a rotary table making it possible to measure arbitrary polarization of the radiation field. The rotary table is connected to a conveyor, which can move in two directions, namely $x$-axis and $y$-axis as shown in Figure 6(a). Besides, the supporter of the horn antenna can be moved forward or backward, getting different value of $l_{distance}$. The flexible cooperation between the rotary table, conveyor, and the supporter of the horn antenna makes it possible to measure the information of the field at arbitrary positions with arbitrary polarization.

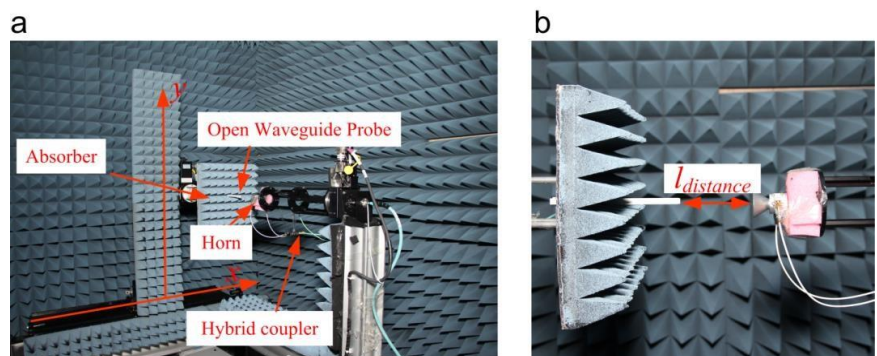

Figure 7 | Schematic of the measurement setup of the horn antenna. (a) Schematic diagram and the coordinate of the measurement system. (b) Photo of the probe and the horn antenna under test.

Supplementary Materials

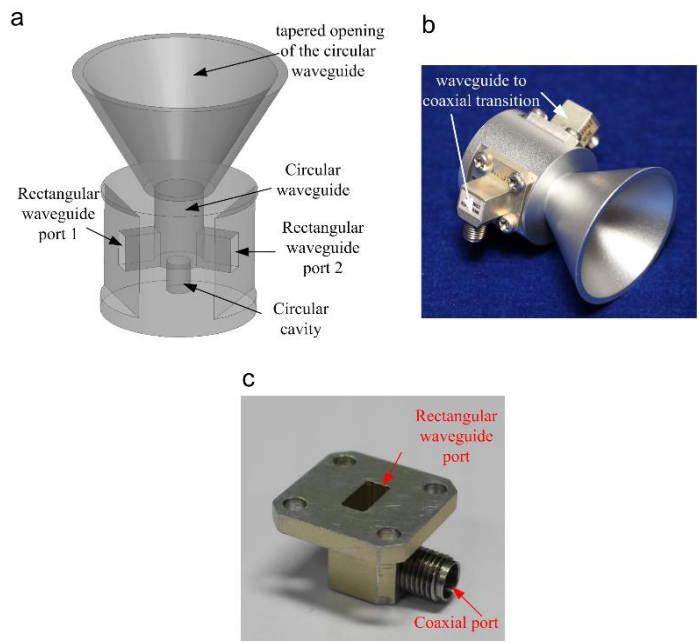

Supplementary Figure 1 | Detailed structures of the horn antenna. (a) Three dimensional perspective of the horn antenna without the feed network. The antenna consists of two rectangular waveguide feeding ports, a circular waveguide and the tapered opening of the circular waveguide. The rectangular waveguide has a standard dimensions operating at Ka-band. Two rectangular waveguides are used as feed of the circular waveguide. When the two ports are excited with property signals, the higher order mode in the circular waveguide can be excited. In this transition between rectangular waveguide and circular waveguide, some higher order modes that we do not want may be excited. To solve this problem, a small circular cavity is added in the end of the circular waveguide. This cavity can dampen the unwanted higher modes. (b) The horn antenna with two rectangular-to-coaxial transitions. In the practice measurement, the antenna is feed with coaxial cable. A rectangular-to-coaxial transition is lunched at each rectangular waveguide port. (c) Prototype of the rectangular-to-coaxial transition

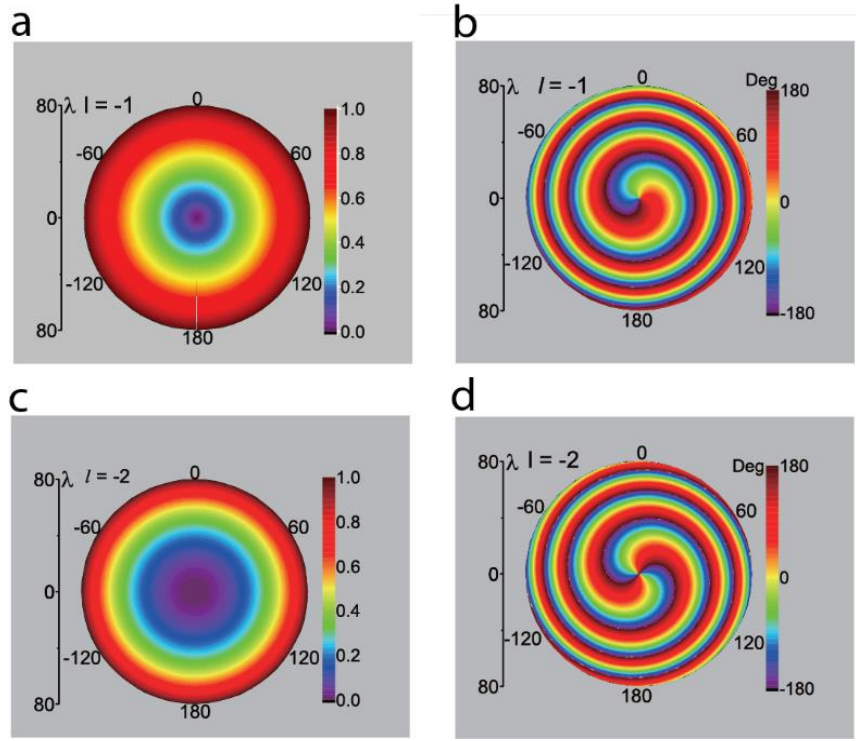

Supplementary Figure 2 | Calculated results of the filed intensity and phase of the field for negative l value. (a) 2-D phase pattern of the twist beams calculated by Matlab with OAM index l = -1, the circular observation window has a radius of ρ=80λ. (b) 2-D amplitude pattern of the twist beams with OAM index l = -1. (c) 2-D phase pattern of the twist beams with OAM index l = -2. (d) 2-D amplitude pattern of the twist beams with OAM index l = -2.

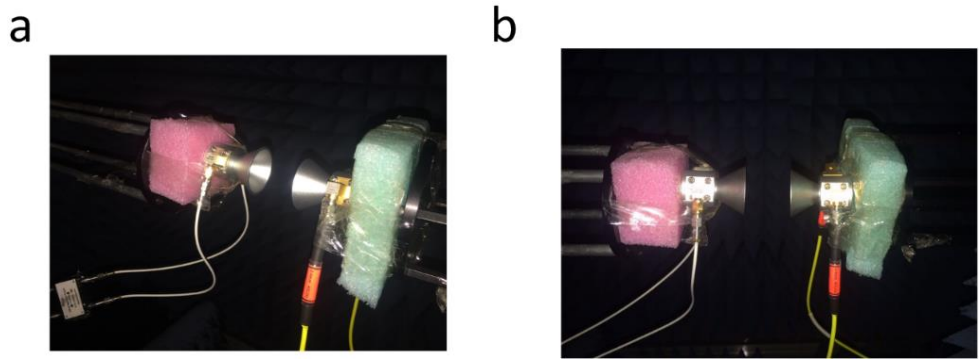

Supplementary Figure 3 | Setup of the Communication link measurement (a) Photo of the two horn antennas placed face to face for measurement. (b) Photo of the two horn antennas placed face to face for measurement (Side view).